\documentclass[12pt,preprint]{aastex}

\newcommand{\etal}{et~al.}

\slugcomment{Accepted for publication in AJ}

\shorttitle{Star Clusters in M33}
\shortauthors{Park and Lee}

\begin{document}

\title{A Catalog of New M33 Star Clusters Based on the HST/WFPC2 Images}

\author{Won-Kee Park and Myung Gyoon Lee}

\affil{Astronomy Program, Department of Physics and Astronomy,
Seoul National University, Seoul 151-742, Korea}
\email{wkpark@astro.snu.ac.kr, mglee@astrog.snu.ac.kr}

\begin{abstract}
We present the result of a survey for star clusters in M33 using the HST/WFPC2 
archive images. 
We have found 104 star clusters, including 32 new ones, in the
images of 24 fields that were not included in the previous
studies. Combining these with previous data in the literature, we increase the
number of M33 star clusters found in the HST images to 242.
We have derived $BVI$ integrated photometry of these star clusters from the CCD
images taken with CFH12k mosaic camera at CFHT.
Integrated color-magnitude diagrams of the M33 star clusters 
are found to be similar in general to those of star clusters in the Large Magellanic Cloud,
except that M33 has a much lower fraction of blue star clusters. 
We find 29 red star clusters with $0.5 \leq (B-V)_0 \leq 1.1$ and $0.7\leq(V-I)_0\leq1.2$, 
which are old globular cluster candidates.
We divide the cluster sample into three groups according to their $(B-V)_0$ color:
blue star clusters with $(B-V)_0 \leq 0.3$, 
intermediate color star clusters with $0.3<(B-V)_0<0.5$, and
red star clusters with $(B-V)_0 \geq 0.5$.
Most of the new clusters in M33
are located along the sequence that is consistent with 
the theoretical evolutionary path
for $Z=0.004$, $Y=0.24$ in the $(B-V)_0$--$(V-I)_0$ diagram, while a few of them are
in the redder side in the $(V-I)_0$ color.
The red clusters are found relatively more 
in the outer region of M33 than the blue and intermediate color clusters, and that
many of the blue stars are located in the HII regions.
The luminosity function  for the blue star clusters shows a peak at 
$M_V \approx -7.3$ mag,  
while that for the intermediate color star cluster  shows a peak at the fainter magnitude $M_V \approx -6.3$ mag. 
The luminosity function for the red star clusters shows also a peak at $M_V \approx -6.8$ mag,
although the number of the clusters is small.

\end{abstract}


\section{Introduction}

Star clusters are an important tool for understanding the formation and evolution of
a galaxy. Being an aggregate of many stars, they are bright enough to be observed at
great distances. 
They span a very broad range of age,
from a few Myrs to more than 10 Gyrs, encapsulating the whole history of a galaxy. 
Age distribution of star clusters in
a galaxy tells us about the formation and destruction history of the clusters and
about the evolution of the galaxy as well.
Spatial distribution of the star clusters delineates the structure of a galaxy,
and their velocity distribution reveals the kinematics of stellar population and the
mass of the host galaxy. 

Recent studies for star clusters in nearby late type galaxies have
revealed details of the star clusters, 
showing that the populations of star clusters are diverse among the galaxies 
(see \citet{lee06} and references therein). 
Local Group galaxies in the nearby universe are ideal targets, due to their
proximity, for studying in detail the properties of star clusters.
There are three spiral galaxies in the Local Group: our Galaxy, M31 and M33.
While star clusters in our Galaxy and M31 have been extensively studied, 
the star clusters in M33 received relatively less attention.

M33 (=NGC 598), the Triangulum galaxy, is a late type spiral galaxy (Scd).
It is located at $\sim 910$ kpc from us \citep{kim02}. 
Its large angular size and inclination of $i = 56\arcdeg $ 
\citep{zar89} make it particularly suitable for studies of its stellar contents.
Its mass, $M \sim 5\times10^{10} M_{\odot}$ \citep{cor03}, falls in the intermediate
range among the Local Group galaxies. 

Surveys for star cluster in M33 have been rather sporadic. Early photometric surveys
by \citet{hil60} and \citet{km60} identified about 25 star cluster candidates.
\citet{md78} reported the discovery of 58 possible star
cluster candidates, including most of the objects studied by \citet{hil60}. 
\citet{cs82} presented the most comprehensive catalog of
250 non-stellar objects in M33, including $BVR$ photometry of 60 star cluster candidates.
Later  \citet{cs88} presented $BVI$ photometry
for 71 additional star cluster candidates.
All these surveys were based on the photographic plates so that they were
limited rather to the outer part of M33.

The first survey for M33 star clusters based on CCD imaging was given by
\citet{moc98} using the $BVI$ images from DIRECT project \citep{kal98, sta98}. 
They presented a list of 51 globular cluster candidates. While they covered the central region
of M33 where previous surveys could not examine in detail, their survey was limited to the
bright star clusters of $V \lesssim 19$ mag, and did not cover the entire area of M33.

The Hubble Space Telescope (HST) offers a great advantage over the ground-based
telescopes for star cluster survey in M33. 
At the distance of M33, star clusters can be easily identified without ambiguity
with HST images. 
Therefore, star cluster samples from HST images are much less likely to be
contaminated by other extended sources such as a background galaxy or an H II region.
\citet{cha99a, cha01}
searched for star clusters in M33 with HST data for the first time.
They reported a discovery of 162 star clusters including 131 previously unknown ones
in the 55 fields of HST/WFPC2 images. Recently, \citet{bed05}
have found 33 star clusters and 51 possible candidates from the $F775W(I)$ image of one
HST/ACS field. 
\citet{sar07a} also reported a discovery of 24 star clusters
including 11 new ones from the $VI$ images of two HST/ACS fields in M33.

However, the areas of M33 covered by the previous HST-based surveys are much smaller
than the entire area of M33. Therefore, it is needed to increase the number of HST fields
to make as complete a sample of star clusters as possible.
The sample of star clusters established from the HST images also serves as a template
for ground-based star cluster surveys.

We searched for new M33 star clusters using the additional HST/WFPC2 images that
became available after \citet{cha01}'s work.
We combined our result with previous M33 star cluster catalogs based on the HST images
\citep{cha99a, cha01, bed05, sar07a}, and obtained homogeneous $BVI$ photometry
for the star clusters from the deep CCD imaging data taken at the Canada-France-Hawaii-Telescope
(CFHT) (Park et al. 2007, in preparation)

In this paper, we present the result of M33 star cluster survey with the HST/WFPC2
images. Throughout this paper, we will refer the star clusters from the catalog of
\citet{bed05} with B designation, and ones from \citet{cha99a, cha01} with CBF.
CS designation means the star clusters from \citet{cs82, cs88}, MKKSS designation
means the ones from \citet{moc98}, MD the ones from \citet{md78}. Finally, SBGHS
designation means the star clusters from \citet{sar07a}. 

The paper is composed as follows.
Section 2 describes the HST/WFPC2 and CFHT data used in this study. 
In Section 3, star
cluster search method and the integrated photometry procedures on CFHT images are
described. Section 4 presents various properties of the star clusters:
Color-magnitude diagrams (CMD), Color-color diagrams (CCD), and spatial distribution of star clusters.
Finally, summary and conclusion are given in the last section. 

\section{Cluster Selection and Photometry}

\subsection{Cluster Selection}
\begin{figure}
\epsscale{1.0}
\plotone{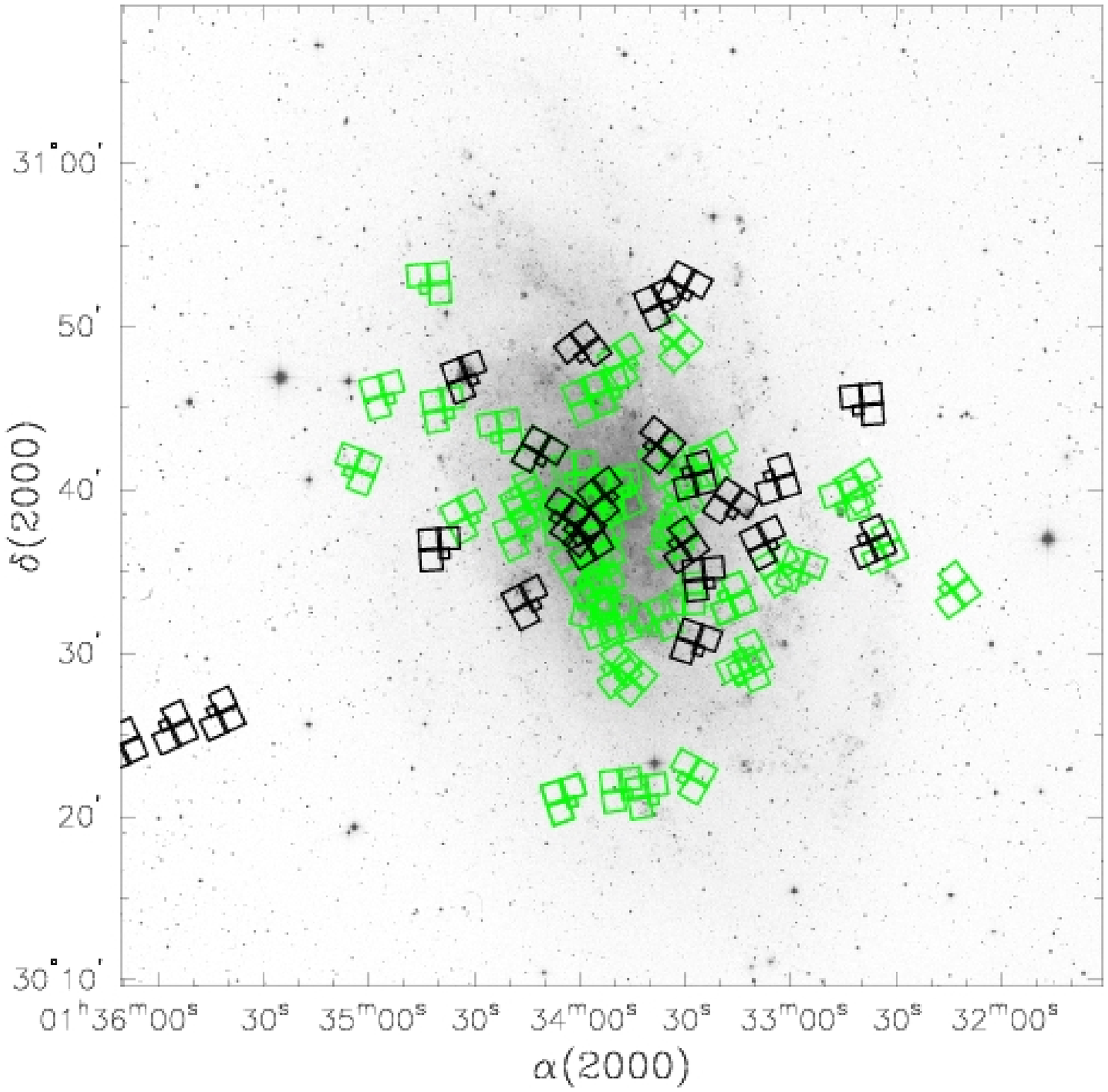}
\caption{ 
Locations of the HST/WFPC2 fields used in this study, overlaid in black on the
grayscale map of $1 \arcdeg \times 1 \arcdeg$ Digitized Sky Survey image of M33.
For comparison, locations of HST/WFPC2 fields for previous star cluster surveys 
\citep{cha99a, cha01} are shown in lighter tone.
\label{fld_location}}
\end{figure}

We used WFPC2 images of M33 in the HST archive that were not used in the
previous studies by \citet{cha99a, cha01}.
The images were obtained using various filter combinations of $F300W$, $F336W(U)$,
$F436W(B)$, $F555W$, $F606W(V)$, and $F814W(I)$. We selected images only with exposure
times longer than 100 sec either in $V$ or $I$. The number of fields thus selected
is 24, as listed in Table~\ref{fld_list}. The locations of the fields are displayed on 
Fig.~\ref{fld_location}, along with the locations of HST/WFPC2 fields for previous star 
cluster surveys \citep{cha99a, cha01}.
Nine of our fields are partially overlapped with previous survey regions.
HST/WFPC2 image sets used by \citet{cha99a, cha01} were also downloaded from the archive, and
images of previously known star clusters on these image sets were utilized as references for
the visual search on the new image sets.

Bright star clusters in M33 are easily resolved on the HST/WFPC2 images so that
we selected star clusters using visual inspection of each image. 
We considered extended objects with the hint of resolved stars as star clusters. 

Since the HST/WFPC2 image sets used in this study have different depths, and they cover 
different environments in M33 with different filters, we could not put a completeness limit 
of the survey for each image. The detection of star
clusters depends not only on the brightness of the star clusters, but also
on their sizes, morphologies, and environments.
Experience from the visual investigation tells us that star clusters with
$V \leq 20$ can be identified in the shallowest images of the
current HST/WFPC2 image sets.

\begin{figure}
\epsscale{1.0}
\plotone{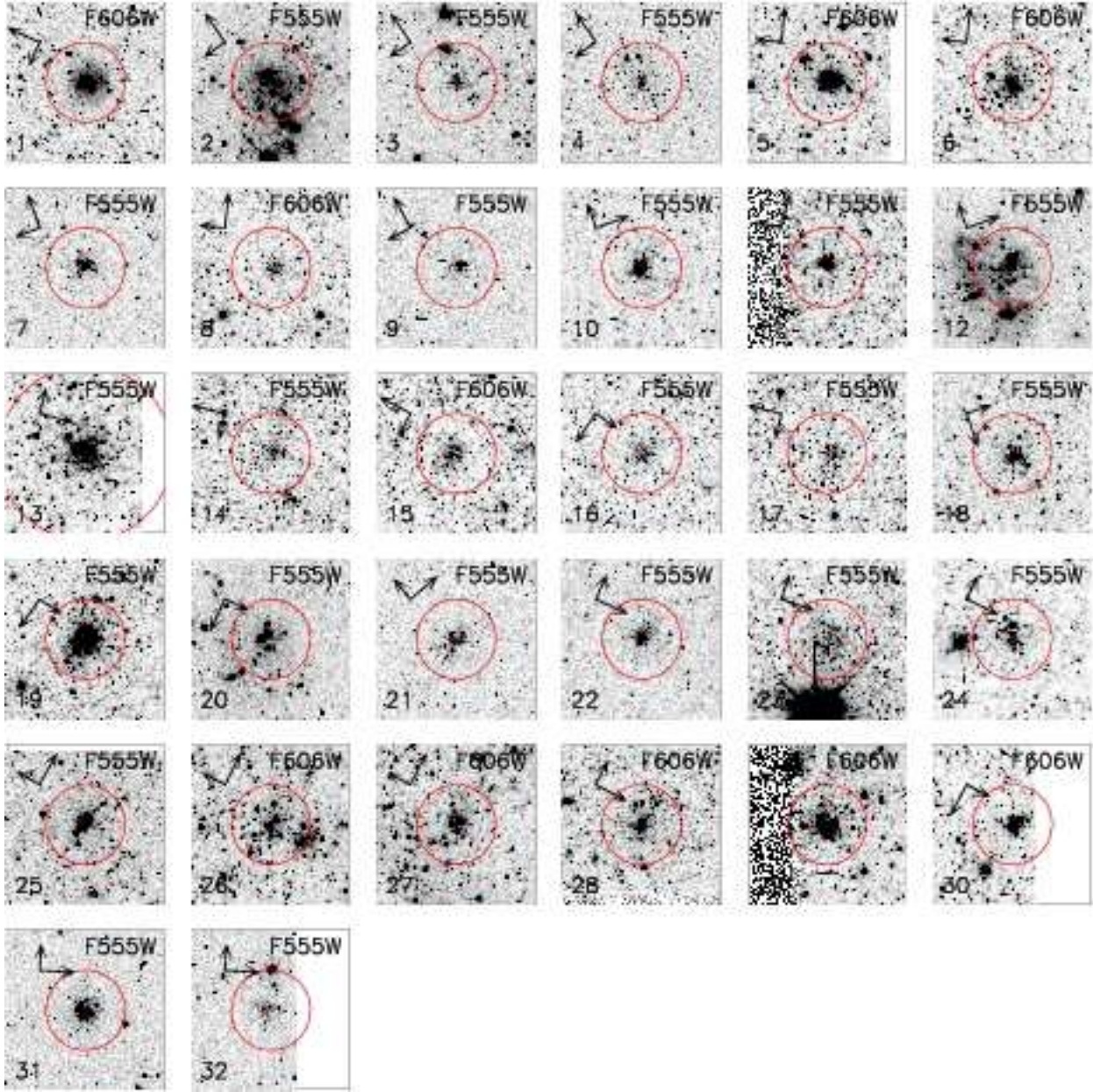}
\caption{ 
A mosaic of the grayscale maps of $V$ band images of new M33 star clusters found in
this study. A circle of $r = 3.0$ arcsec 
is drawn around the cluster to show the scale of each image. 
Longer and shorter arrows indicate, respectively, north and east.
The labels in the right lower corner of each image represent the ID of each star cluster,
and the ones in upper right corner the filter name of the each image.
\label{newcl}}
\end{figure}

From the investigation of HST/WFPC2 fields, we have found 104 star clusters,
32 of which are new ones.
The 32 new star clusters are listed in Table~\ref{newcl_list}, 
and their $V$-band grayscale images are shown in Fig.~\ref{newcl}.
During the investigation, we identified eight CBF star clusters that were
redundantly identified and registered under two different IDs: CBF-15 (CBF-45),
CBF-20 (CBF-126), CBF-22 (CBF-91), CBF-53 (CBF-87), CBF-54 (CBF-85), CBF-56
(CBF-156), CBF-59 (CBF-95), and CBF-60 (CBF-94). The two star clusters in each
pair have not only similar coordinates, but also similar $V$ magnitudes with
differences $<0.2$ mag.
We inspected carefully the HST/WFPC2 images of the regions where eight pairs
are located, and we found neither two star clusters of similar magnitudes in
the regions, nor any possible substructures (i.e. multiple OB groups in a 
cluster) in the clusters that \citet{cha99a, cha01} might have recognized as 
two separate star clusters. Therefore, we regard these eight cases as 
redundant identifications of the same star cluster.
Throughout this paper, we will present the total number of CBF
samples as 154 instead of 162. From these 154 CBF samples, we identified 45 star
clusters on our image sets.

One of our fields partially overlaps the region where \citet{bed05} searched for
star clusters with an HST/ACS image, where they found 33 star clusters and 51 star
cluster candidates. 14 star clusters and 18 star cluster candidates fall on
our HST/WFPC2 image. We could identify 13 star clusters except cluster ID 15 in \citet{bed05}'s
list that seems to be too faint to be seen on our $F606W$ image. We also inspected the 18 star
cluster candidates on our image. However, they are so compact in size that it was impossible
to tell whether they are star clusters, except for cluster ID 83 in \citet{bed05}'s list.
We classified this one as a star cluster after careful inspection of the image.

In addition, we found that 12 star cluster candidates from previous ground-based surveys
\citep{cs82, moc98, md78}
are real star clusters. The coordinates, photometric properties, and cross identification
among the various studies for these 71 previously known clusters are listed in Table~\ref{othercl_list}.

We make a catalog of 242 star clusters in M33, combining our result with the previous 
results based on HST-based surveys \citep{cha99a,cha01, bed05, sar07a}.
It is the largest catalog of M33 star clusters available to date which includes only
the star clusters confirmed in the HST images. 
It consists of 32 newly found star clusters in this study,
12 previous star cluster candidates that were confirmed to be the genuine ones in this
study, 154 star clusters from \citet{cha99a, cha01}, 32 star clusters
out of 33 star clusters in \citet{bed05}, one star cluster from the star
cluster candidate list of \citet{bed05}, and 11 new star clusters from the list of
\citet{sar07a}. 
The positions and the photometry of all M33 star clusters not
included in Tables~\ref{newcl_list} and \ref{othercl_list} are listed in Table~\ref{prevcl_list}.

\subsection{Photometry}

All star clusters in the combined catalog were observed with HST in different
conditions with different purpose. The combination of the filters used for the
HST observation varies depending on the field.
To secure homogeneous photometry data for all clusters,
we used the integrated photometry of these clusters derived
from the ground-based images given in Park et al. (2007, in preparation)
instead of deriving the photometry from the HST images.

Park et al. (2007) presented $BVI$ integrated aperture photometry of clusters and cluster
candidates in the $50\arcmin \times 80\arcmin$ field of M33 
based on the CCD images taken with
CFH12k mosaic camera at the Canada-France-Hawaii Telescope (CFHT).
The typical seeing of the data is $\sim 0.9$ arcsec for all filters, and the
point sources of $V\sim23$ can be measured with errors smaller than $0.1$.
Details of the photometry are given in Park et al. (2007), and  
we give a brief description here. 

Examination of an aperture growth curve of isolated bright clusters 
on  CFH12k CCD images indicates that most
of star cluster light is contained within the aperture of $r \approx 4$ arcsec.
However, using such an aperture would cause a large scatter in the measurement of some clusters,
because star clusters are often located in the crowded field
in M33 so that some neighbor objects happen to be inside the same aperture.
For this reason, we used an aperture of $r=2$ arcsec for the measurement of color, although
the aperture of $r = 4$ arcsec was used for $V$ magnitude measurement to measure the total light of 
the star clusters.

Some of the star clusters we missed happened to fall on the gaps between the chips in mosaic
camera, and some of them happen to be located in so crowded a region that we could not
derive their photometry from the CFHT images.

To supplement the photometry data for star clusters that could not be measured with CFH12k
data, aperture photometry was carried out on the HST/WFPC2 images.
We made photometry on the fields that were observed with more than two filters since
instrumental magnitude cannot be converted to standard system without color information.
For most of fields, two HST/WFPC2 images of identical pointings exist per filter.
Fields observed with only one exposure were not included in the measurement.
Two images were combined with COMBINE task in IRAF\footnote{IRAF is distributed by the
National Optical Astronomy Observatories, which are operated by the Association of Universities
for Research in Astronomy, Inc., under cooperative agreement with the National Science
Foundation} with CRREJECT option to remove the cosmic rays.
Cosmetic defects were masked with data quality images that accompany the scientific
images.
The centers of the clusters were determined visually with IRAF/IMEXAMINE task. Then
IRAF/PHOT task was used to measure the instrumental magnitudes of star clusters on the
combined images.
To be consistent with previous HST-based studies \citep{cha99a, cha01, sar07a}, 
aperture of $r=2.2$ arcsec was adopted for the magnitude measurement, while aperture of
$r=1.0$ arcsec for color measurement. Local median values were measured as sky background
level in the annulus of inner radius of $r=3.5$ arcsec and outer radius of $r=5.0$ arcsec
around each star cluster.
The CTE correction was applied to the instrumental magnitudes according to the prescription
given at Dolphin (2002)'s web site\footnote{\url{http://purcell.as.arizona.edu/wfpc2\_calib/}}.
However, we did not correct for the geometrical distortions in the WFPC2 CCD. The instrumental
magnitudes were converted to standard system using the relation also found at the same web site
iteratively. 
In the end, we derived the photometry of 227 out of 242 star clusters with both CFH12k
and HST/WFPC2 images.

We compared our photometry of the clusters with previous photometry.
There are 154 clusters in common with \citet{cha99a, cha01}, 
16 clusters in common with \citet{cs82}, 
12 clusters in common with \citet{moc98}, 
and 11 clusters in common with \citet{sar07a}. 
Note that all photometry was carried
out with different aperture sizes: 
\citet{cha99a, cha01} and \citet{sar07a} used nominal aperture
of $r = 2.2$ arcsec for magnitude measurement, 
while \citet{cs82, cs88} used apertures
of $r=2.4$ and $r=3.7$ arcsec, respectively. 
\citet{moc98} used an aperture of $r=2.88$ arcsec.
However, aperture sizes in all cases were determined to contain as much as possible the
light from a star cluster, so the
resultant aperture magnitudes for a star cluster should be more or less the same. 

\begin{figure}
\epsscale{1.00}
\plotone{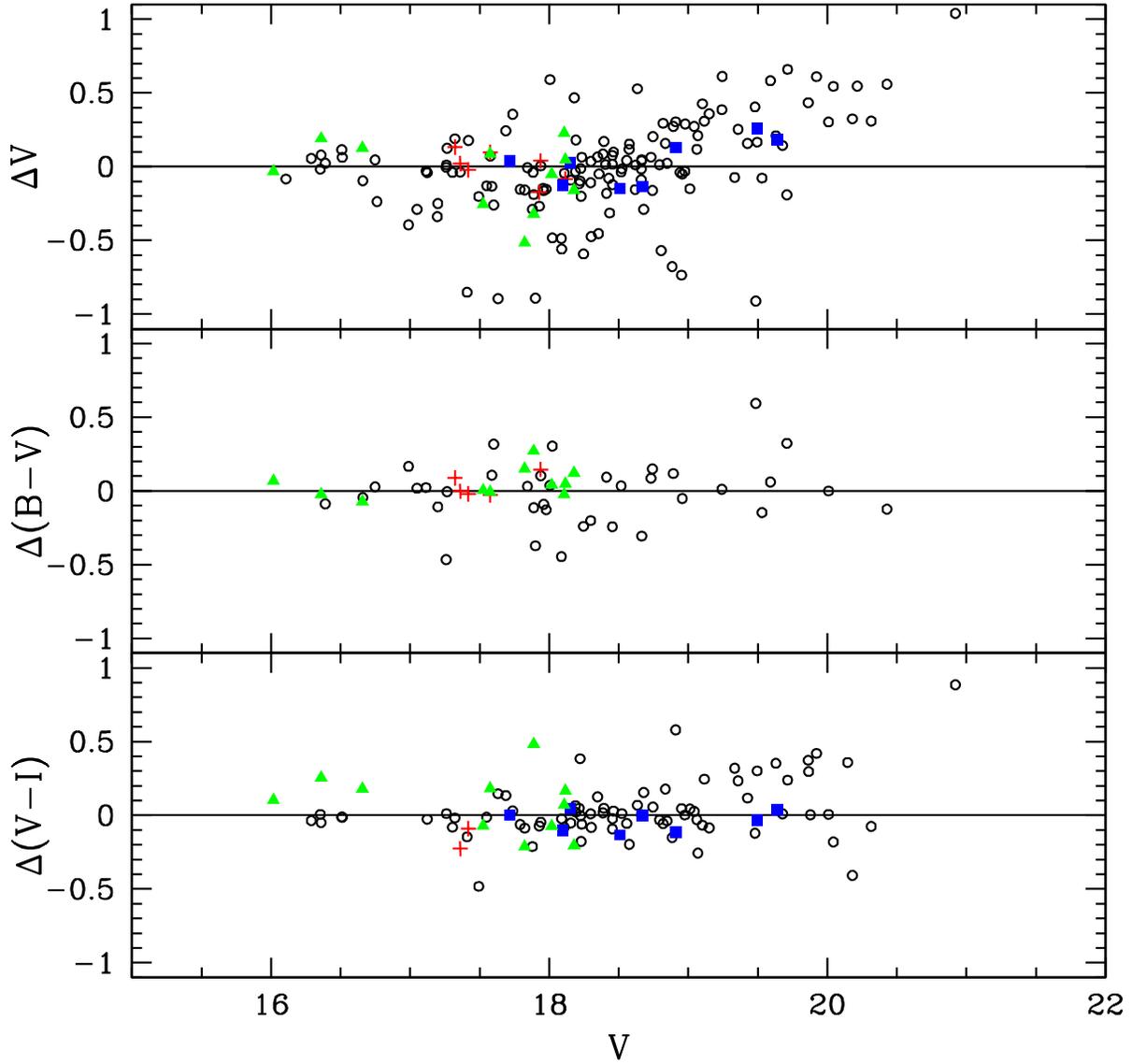} 
\caption{ 
Comparison of the photometry between this study and previous studies.
Open circles represent the photometry difference of $\Delta$(this study - \citet{cha99a, cha01}),
filled squares the difference of $\Delta$(this study - \citet{sar07a}), filled triangles 
$\Delta$(this study - \citet{moc98}), and crosses denotes the difference of 
$\Delta$(this study - \citet{cs82, cs88}).
\label{photdiff}}
\end{figure}

Fig.~\ref{photdiff} shows the comparison of our photometry of the clusters 
with previous photometry.
Large scatter made it difficult to see the difference in magnitude, but for bright
star clusters ($V \leq 19$), there is little magnitude differences between this
and other studies:
$\Delta V$(this study--\citet{cha99a, cha01}) $= -0.04$ with $\sigma = 0.19$, 
$\Delta V$(this study--\citet{moc98}) $= -0.04$ with $\sigma = 0.17$,
$\Delta V$(this study--\citet{cs82, cs88}) $= 0.00$ with $\sigma = 0.10$, and
$\Delta V$(this study--\citet{sar07a}) $= -0.04$ with $\sigma = 0.12$, respectively.
For fainter sources, however, our measurements get systematically fainter than
the other space based measurements.
Given that most of the star clusters are located in the region near the M33 center
where crowding is severe and sky varies significantly, the sky background level
was likely to be overestimated in ground-based images due to the unresolved faint
stars, which would lead to a fainter estimation of source brightness.

Both the $(B-V)$ and $(V-I)$ colors show good agreement with all the previous studies 
down to the faintest magnitude in our samples. The differences between this study
and other photometry are  
$\Delta(B-V)$(this study--\citet{cha99a, cha01})$= -0.05$ with $\sigma = 0.20$,
$\Delta(V-I)$(this study--\citet{cha99a, cha01})$= -0.01$ with $\sigma = 0.10$,
$\Delta(B-V)$(this study--\citet{moc98})$= 0.06$ with $\sigma = 0.10$,
$\Delta(V-I)$(this study--\citet{moc98})$= 0.08$ with $\sigma = 0.22$,
$\Delta(B-V)$(this study--\citet{cs82, cs88})$= 0.04 $ with $\sigma = 0.08$,
and $\Delta(V-I)$(this study--\citet{sar07a})$ = -0.04 $ with $\sigma = 0.08$. 
The only exception to this tendency is the $(V-I)$ difference between ours and those of
\citet{cs82, cs88}, which turned out to be :
$\Delta(V-I)$(this study--\citet{cs82, cs88}) $= -0.16 $ with $\sigma = 0.10$.

\subsection{Reddening}

To get the intrinsic color of the star clusters, the photometric measurements
must be corrected for reddening.
We estimated the foreground reddening for each star cluster using the $(U-B)$--$(B-V)$ diagram
for early type stars in the field covering the cluster.
We used the $UBVRI$ photometry of stars in the area of M33 covered in the Local 
Group Survey given by \citet{mas06}. The $E(B-V)$ derived for each cluster is listed in
the fourth column of Table~\ref{newcl_list}, \ref{othercl_list}, and \ref{prevcl_list}.

$E(B-V)$ values we derived for M33 star clusters range from $\sim 0.05$ to $\sim 0.20$.
For about 130 star clusters, the value of $E(B-V)$ turned
out to be $E(B-V) = 0.1$, and for about 100 star clusters the values of $E(B-V)$ were found
to be $E(B-V) \geq 0.15$.
\citet{cha99b} obtained $E(B-V)$ toward their 60 star clusters either from the
photometry of neighboring individual stars by applying reddening free parameters or by using
$E(B-V)$ values from \citet{mas95}. The $E(B-V)$ values range $0.06\sim0.33$, and most of them
are found to be $0.10$. \citet{sar00} also derived reddening for their
10 globular cluster samples from the shapes of the red giant branches (RGB) in the
color-magnitude diagrams of resolved stars. Their $E(V-I)$ estimates are below $0.1$, except
for $E(V-I)=0.25\pm0.03$ for Cluster ID 131 (=CS-H10). \citet{sar07a} also obtained $E(V-I)$
for 17 star clusters in the same way as \citet{sar00}, which $E(V-I)$ ranges from $0.10
\sim 0.45$.

Since our error in $E(B-V)$ measurement is as large as $\sim 0.05$, it is quite difficult
to see the consistency between our result and other results. But the fact that $E(B-V)$
distribution shows peak at $E(B-V) \sim 0.10$, a typical value found in previous studies
\citep{cha99b}, indicated that the result obtained in this study is in general consistent
with previous studies.

\section{Result}

\subsection{Color-magnitude diagram}

\begin{figure}
\epsscale{0.8}
\plotone{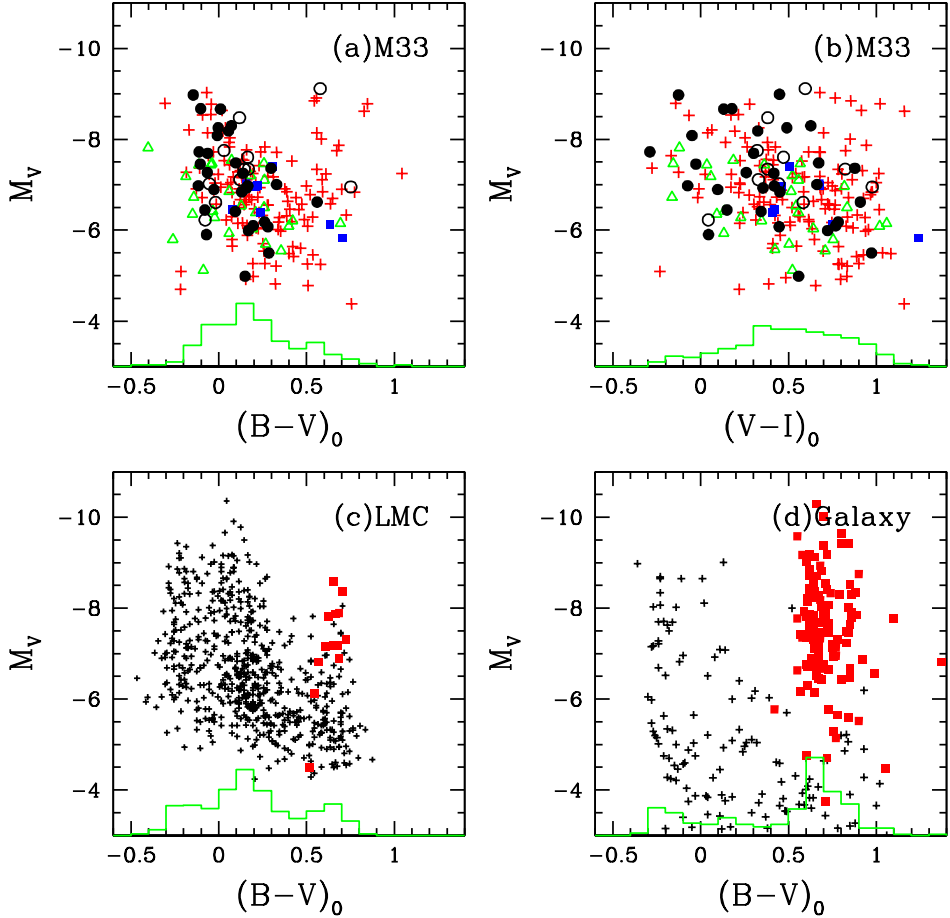} 
\caption{ 
Integrated color-magnitude diagrams (CMD) of the star clusters in M33, Large
Magellanic Cloud (LMC), and our Galaxy. Panels (a) and (b) show the $M_V$--$(B-V)_0$
CMD and the $M_V$--$(V-I)_0$ CMD of M33 star clusters, respectively.
In both panels, filled circles denote the star clusters newly
discovered in this study, and open circles denote the previously known star
clusters that were confirmed in this study. Filled squares denote the star
clusters from \citet{sar07a}, Crosses the ones in the list of \citet{cha99a, cha01},
while triangles the star clusters from \citet{bed05}. Panel (c): integrated
$M_V$--$(B-V)_0$ CMD of star clusters in the LMC \citep{bic96}.
Panel (d): the integrated $M_V$--$(B-V)_0$ CMD of open clusters \citep{lat02} and 
globular clusters \citep{har96} in our Galaxy. 
Globular clusters are marked with filled squares in both panels. Histograms in all
panels show the normalized distributions of colors for star cluster samples plotted
in each panel.
\label{cmd}}
\end{figure}

Fig.~\ref{cmd} displays the integrated $M_V - (B-V)_0$ and $M_V - (V-I)_0$ 
color-magnitude diagrams (CMD) of the star clusters in M33. 
Absolute magnitudes of the star clusters were derived for the adopted 
distance modulus of $(m - M)_0 = 24.8$ \citep{kim02} and 
 the visual extinction to reddening ratio $R_V= 3.31$ \citep{sch98}.
For comparison, we display the CMD for the star clusters in the LMC
\citep{bic96} in the panel (c), 
and the CMDs for the open clusters \citep{lat02} and globular clusters
\citep{har96} in our Galaxy in panel (d). 
We adopted 
a value for the foreground reddening of $E(B-V) = 0.075$ \citep{sch98}
and the distance modulus of $(m - M)_0 = 18.5$ for the LMC clusters.

Several features are noted in Fig.~\ref{cmd}.
First, the star clusters in all three galaxies are roughly separated into blue and red
groups with a color boundary of $(B-V)_0 \simeq 0.5$ in the $M_V - (B-V)_0$ CMD. However, 
this separation for the M33 clusters in the $M_V - (V-I)_0$ CMD is not 
as clear as in the $M_V - (B-V)_0$ CMD. 
Second, the dominant population in M33 is blue clusters, as known before \citep{cs82, cha99b}. 
Most of the new clusters found or the clusters confirmed in this study are also blue clusters.
Third, 
the brightest blue clusters in M33 are as bright as $M_V \approx -9.0$ mag, which is similar
to the value for the Galactic open clusters, but about 1.5 magnitude fainter than
that for the LMC.
Fourth, the blue envelope for the blue clusters in M33 in the $M_V$--$(B-V)_0$ CMD is slightly
redder than that for the LMC blue clusters, as seen from the normalized $(B-V)_0$ color
histograms in the panel (a) and (c).
Fifth, there are only a small number of red clusters with $(B-V)_0>0.5$ in M33.
The $(B-V)_0$ color range of the red clusters in M33, $0.5<(B-V)_0<1.1$,
is similar to that for the Galactic globular clusters, but is much larger than
that for the LMC red clusters, $0.5<(B-V)_0<0.8$.
We found only one new red cluster, and confirmed two known red clusters. 
Cluster ID 85 (ID 27 in \citet{moc98}), one of the two red clusters confirmed in this study,
happens to be the brightest ($M_V=-9.070$ and $(B-V)_0=0.575$) among the known red clusters
in M33. The reddest cluster is cluster ID 89, which has integrated colors 
$(B-V)_0 = 1.579$, and $(V-I)_0 = 1.810$.

It is interesting to note that there are only a few red star clusters that are fainter
than $M_V \sim -6.0$ in M33,  
while there are many such faint red star clusters seen in the LMC. 
The lack of faint red clusters in our catalog of M33 clusters 
may be due to incompleteness of our cluster search 
or may be an intrinsic feature of the M33 star cluster system. 
While we selected HST/WFPC2 images on the basis of $V$ or $I$ exposure
times, many of resultant image sets were obtained with combinations of $V$
and shorter wavelength passbands such as $U$ and $B$, 
rather than the combinations of $V$ and $I$. 
So there is a possibility that our star cluster detection might have been biased 
toward blue star clusters. 
However, it may be an intrinsic feature, considering that \citet{cha01} already
showed that M33 red star clusters with $(V-I)_0 > 0.78$ has 
log normal type luminosity function (LF), 
while intermediate color clusters with $(V-I)_0 \leq 0.78$ show no turnover down to
the completeness limit in their LF.

\subsection{Color-color diagram}

\begin{figure}
\epsscale{0.8}
\plotone{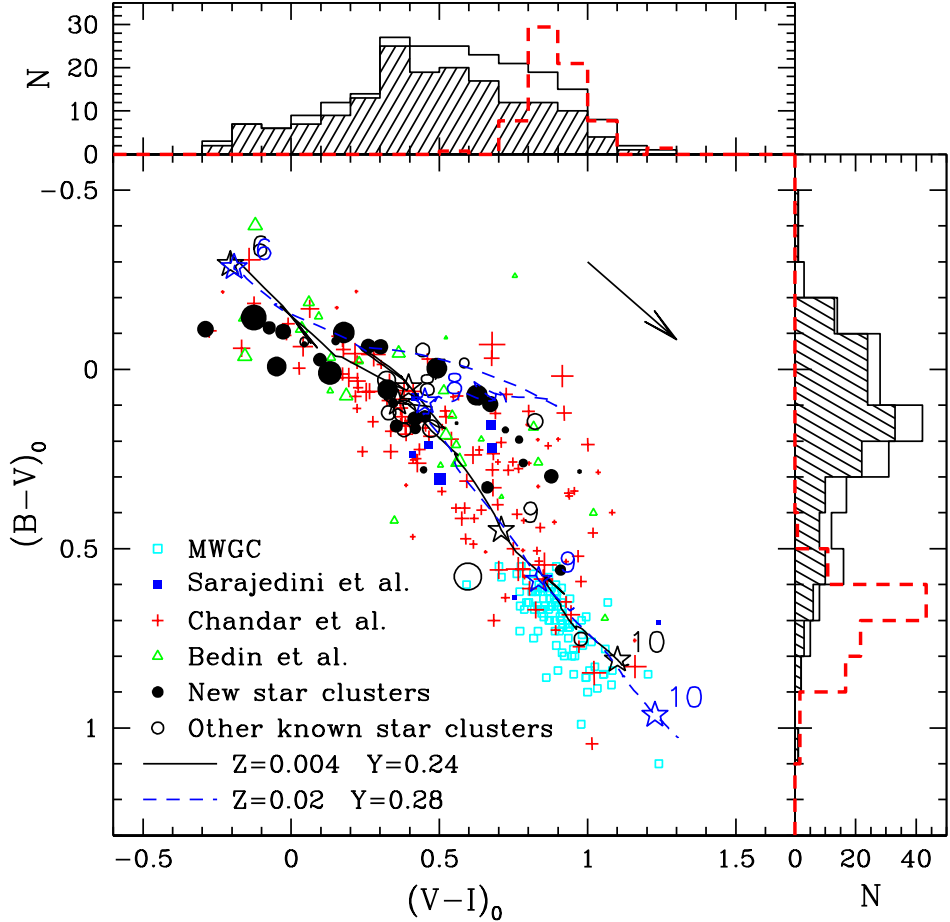} 
\caption{ 
$(B-V)_0$ -- $(V-I)_0$ diagram of the star clusters in M33. 
Same symbols are used for each group of star clusters as in panels (a) and (b) of 
Fig.~\ref{cmd}. The sizes of symbols for M33 star clusters represent the relative
brightness of the star cluster: the bigger, the brighter.
Theoretical evolutionary paths from the Single Stellar Population(SSP) model \citep{bru03} 
for ${\rm Z}=0.004$, ${\rm Y}=0.24$ (solid line), 
${\rm Z}=0.02$, and ${\rm Y}=0.28$ (dashed line) are
drawn with star symbols for every dex in age from $10^6$ years to $10^{10}$ years.
The arrow represents the reddening direction.
Open histograms at upper and right side of the figure represent the distributions of the
integrated colors for all M33 star clusters, while hatched histograms for star clusters brighter
than $V \leq 19.0$. Color distributions of Galactic globular clusters are represented with
dashed lines. Scales of Galactic globular cluster histograms are adjusted to fit in the
figures.
\label{ccd}}
\end{figure}

Fig.~\ref{ccd} shows the integrated $(B-V)_0$--$(V-I)_0$ color-color diagram (CCD)
of M33 star clusters, along with $(B-V)_0$ and $(V-I)_0$ histograms of M33 star clusters.
Galactic globular clusters are also plotted for comparison.
We overlaid the theoretical evolutionary path for the Single Stellar Population (SSP)
\citep{bru03} for $Z=0.004$, $Y=0.24$ that was reported to give the best overall fit to
their M33 star cluster samples which younger than  $log(t) \leq 9.0$ \citep{cha99b}.
For comparison, evolutionary path of SSP for $Z=0.02$, $Y=0.28$ is also overlaid.
  
Most of the new clusters in M33 including the ones from \citet{bed05}
are located along the sequence that is consistent with 
the theoretical evolutionary path 
for $Z=0.004$, $Y=0.24$, while a few of them are in the redder side in the
$(V-I)_0$ color.
This is similar to the case of previously known clusters in M33 and also
to the case of the star clusters in M51 \citep{hwa07}.
Some bright clusters among these redder clusters are consistent with the 
theoretical evolutionary path  for $Z=0.02$, $Y=0.28$. However, the rest of them
cannot be explained with any other models.
The continuous distribution of star clusters
along the model line indicates that M33 star clusters have been formed continuously from
the epoch of the first star cluster formation until recent time. 

The $(V-I)_0$ distribution
shows only one peak at $(V-I)_0 = 0.45$ and monotonically decrease
toward red end, while the $(B-V)_0$ distribution shows two distinct peaks
at $(B-V)_0 \sim 0.15$ and $(B-V)_0 \sim 0.55$.
The $(B-V)_0$ distribution is similar to that given by \citet{sar07b},
except that the red peak at $(B-V)_0 \sim 0.55$ in our result, is 0.1 bluer than 
their result, $(B-V)_0 \sim 0.65$.
The blue peak corresponds to the age of $\sim 2\times10^8$ year and the red one to the age
of $\sim 1\times 10^9$ year, if we assume the metallicity of the star clusters, $Z=0.004$.

Based on the integrated photometry of the clusters,
we can select old globular cluster candidates from our catalog. The $(B-V)_0$ color
distribution shows the blue end of red star cluster population is located at $(B-V)_0 \sim 0.5$,
which is similar to that of Galactic globular clusters. 
Although we do not see any similar red peak at $(V-I)_0$ color distribution of M33 star clusters,
it is reasonable to set the color criterion for $(V-I)_0$ color as well, in accordance with
the $(V-I)_0$ distribution of Galactic globular clusters.
The color ranges for most of Galactic globular clusters in Fig.~\ref{ccd} are
$0.5 \leq (B-V)_0 \leq 1.1$ and $0.7 \leq (V-I)_0 \leq 1.2$.
There are 29 star clusters with this color range 
in our catalog, which are old globular cluster candidates in M33. Ten of them are
newly found in this study. The globular cluster candidates
are listed in Table.~\ref{gccand_list}, along with their integrated photometry and
cross identifications with previous catalogs.

Nine out of ten globular clusters in \citet{sar00} satisfy our color selection
criterion.
One globular cluster, C38 in \citet{cs82} (=CBF-114), is not included because its color
$(V-I)_0 = 0.684$ is a little bluer than the blue limit of the criterion.
Our list of 29 globular cluster candidates also includes three oldest star clusters in the
list of \citet{sar07a}, clusters ID 58, 131, and 228
that are all estimated to be older than $log(t) \sim 9.8$. 

\subsection{Spatial distribution}

\begin{figure}
\epsscale{1.00}
\plotone{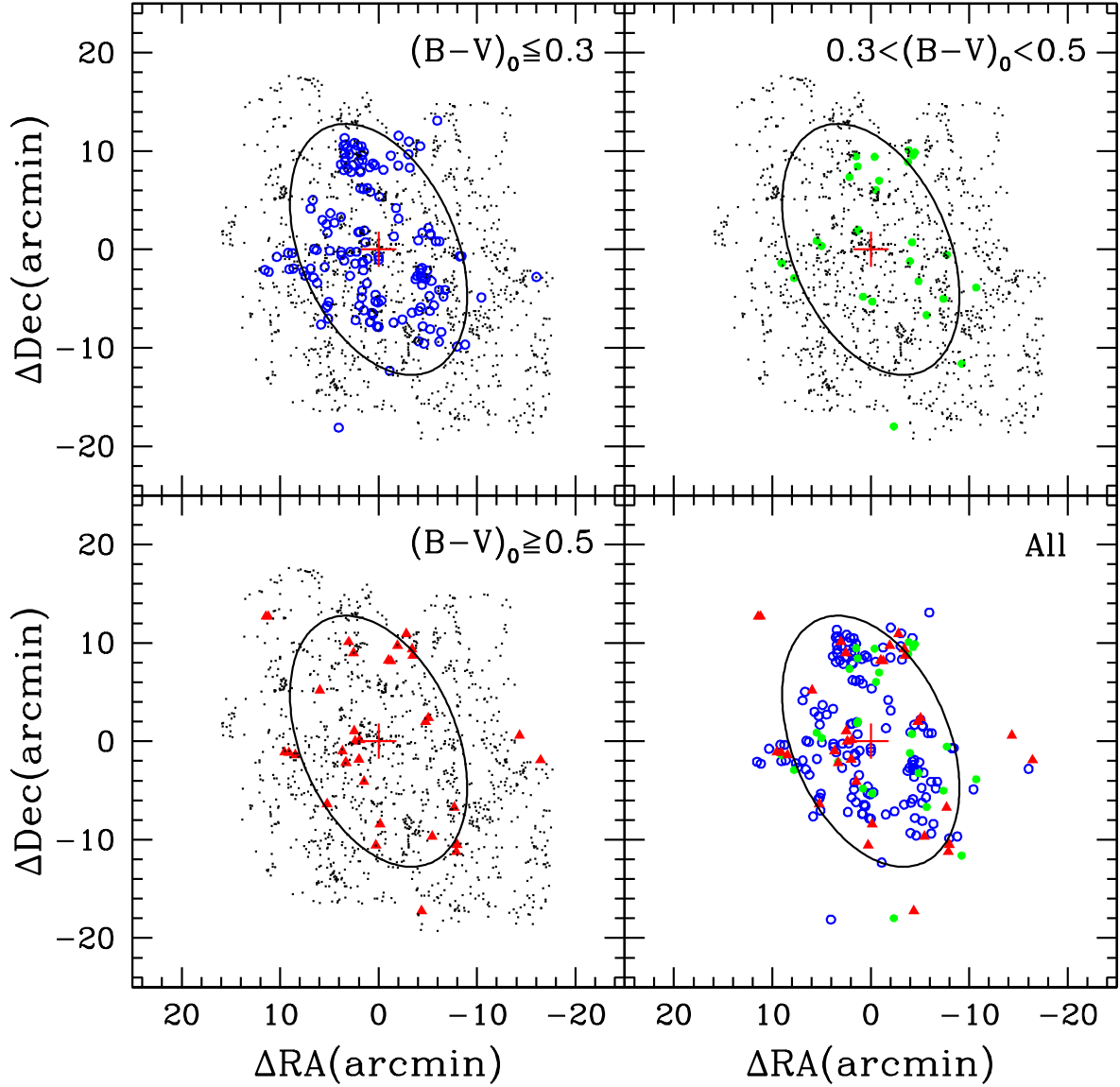} 
\caption{
Spatial distribution of M33 star clusters: (a) blue star clusters with $(B-V)_0 \leq 0.3$,
(b) intermediate color star clusters with $0.3<(B-V)_0<0.5$,
(c) red star clusters with $(B-V)_0 \geq 0.5$, and (d) all star clusters in our catalog.
Small dots represent the locations of H II regions in M33 \citep{hod99}.
The center of M33 is marked with big plus sign. 
Schematic representation of M33
is shown on each panel by an ellipse with major axis length corresponding
to the effective aperture($A_e$) of M33.
\label{spatial_dist}}
\end{figure}

We divided the M33 star cluster sample into three groups according to their $(B-V)_0$ color
for further analysis: blue star clusters with $(B-V)_0\leq0.3$ ,
 intermediate color star clusters with $0.3<(B-V)_0<0.5$, 
and red star clusters with $(B-V)_0\geq0.5$.
Fig.~\ref{spatial_dist} displays the spatial distribution of each color group. 
We also plotted for comparison the locations of H II regions \citep{hod99} 
that trace well the spiral arm structures of M33.
Since HST/WFPC2 observations covered in our catalog do not cover the entire area of M33, 
the spatial distribution of star clusters cannot be used for the study of spatial 
structure of the M33 cluster system.
Fig.~\ref{spatial_dist} shows that the red clusters are found relatively more 
in the outer region of M33 than the blue and intermediate color clusters, and that
many of the blue stars are located in the HII regions.

Note the regions investigated in this study include three fields far east from M33 center
(See Fig.~\ref{fld_location} for locations of these fields.). Had any star clusters been
found in these regions, it would give a clue on the extent of star cluster system
in M33. However, no star clusters were found in these regions with our image sets.

\subsection{Cluster luminosity function}

\begin{figure}
\epsscale{1.00}
\plotone{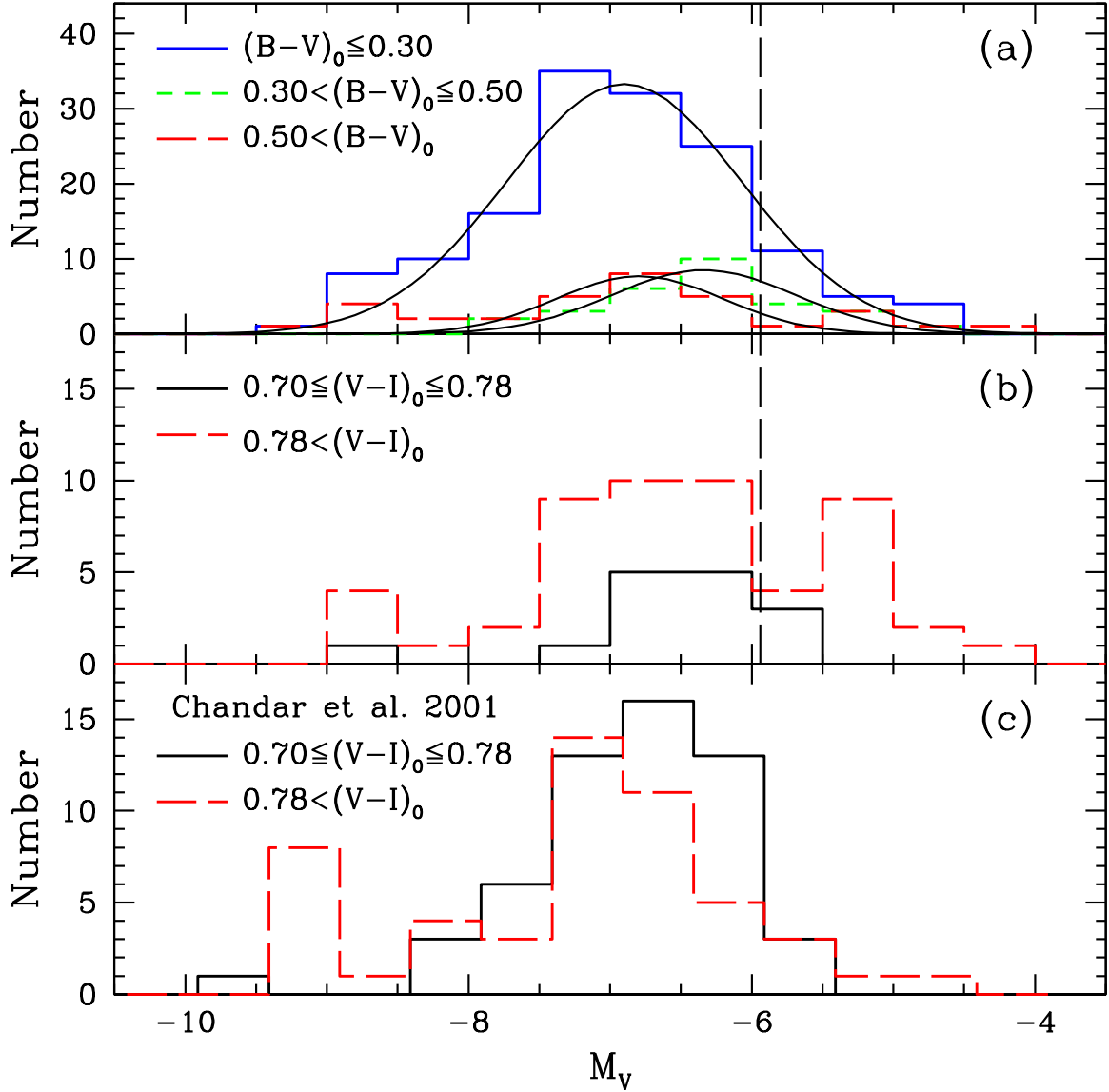} 
\caption{
Luminosity function (LF) of M33 star clusters. (a) LFs 
for blue, intermediate color, and red star clusters
that were classified according to their $(B-V)_0$ color.
(b) LFs for intermediate color clusters and red clusters 
that were selected according to their $(V-I)_0$ color as used in \citet{cha01}.
(c) LFs for intermediate color clusters and red clusters derived by \citet{cha01},
adjusted according to the distance modulus adopted in this study.
Vertical dashed lines indicate roughly the cluster survey limit of this study.
\label{lf}}
\end{figure}

With the largest sample of star clusters available, we derived the luminosity
function (LF) of star clusters, displaying it in Fig.~\ref{lf}(a).
Although we did not estimate the completeness of our survey, it appears that  our survey is
reasonably complete for clusters brighter than $V \leq 19$. Therefore, only the bright part of
LF is considered in further discussion. 

In Fig.~\ref{lf}(a) the LF for the blue star clusters shows a peak at $M_V \approx -7.3$ mag, 
and decreases as the magnitude gets fainter, 
while the intermediate color star cluster LF shows a peak at the fainter magnitude $M_V \approx -6.3$ mag. 
The LF for the red star clusters shows also a peak at
$M_V \approx -6.8$ mag, although the number of the clusters is small.
 
We fitted the LFs for $V<19$ mag using a single Gaussian function, 
obtaining 
$M_V(\textrm{peak}) = -6.90\pm0.04$ mag and $\sigma (M_V ) = 1.96$ for the blue star clusters, 
$M_V(\textrm{peak}) = -6.34\pm0.05$ mag and $\sigma (M_V ) = 1.53$ for the intermediate color star clusters, and 
$M_V(\textrm{peak}) = -6.79\pm0.06$ mag and $\sigma (M_V ) = 1.39$ for the red star clusters.

\citet{cha01} presented the LF of the M33 star clusters for two color groups:
$0.70<(V-I)_0<0.78$ (intermediate color star clusters) 
and $0.78<(V-I)_0$ (red  star clusters). 
So we derived the cluster LFs for the same color
groups, as shown in Fig.~\ref{lf}(b). We also plotted the cluster LFs given by \citet{cha01} 
in Fig.~\ref{lf}(c) for comparison. We adjusted the LFs given by \citet{cha01} 
applying the same distance modulus for M33 (we adopted 
$(m-M)_0= 24.80$, while \citet{cha01} adopted $(m-M)_0= 24.64$),
shifting the magnitude bins of \citet{cha01}'s LF 0.16 magnitude brighter.

In Fig.~\ref{lf}(b) the LFs for the intermediate color and red star clusters derived
according to the $(V-I)_0$ color show peaks at 
$M_V \approx -6.8$ mag, which is roughly similar to those 
derived according to the $(B-V)_0$ color in Fig.~\ref{lf}(a). 
However, the LFs for the intermediate color and red star clusters derived in this study are
somewhat different from those given by \citet{cha01} in Fig.~\ref{lf}(c).
Our LFs contain a larger number of faint star clusters than those given by \citet{cha01}.
The peak magnitude in the LF for the red star clusters derived 
in this study is about 0.3 magnitude fainter 
than that of \citet{cha01}, $M_V \approx -7.1$ mag, while 
the peak magnitude in the LF for the intermediate color star clusters 
in this study is similar to that of \citet{cha01}'s, $M_V \approx -6.7$ mag.

\citet{cha01} sampled brighter star clusters more
than in our survey. \citet{cha01} combined their catalog based
on HST/WFPC2 data with the previous catalogs of M33 star clusters 
to increase the number of the sample for deriving the LFs. 
We used only star cluster samples obtained with HST observation. 
Previous ground based surveys covered a wider area in M33 with shallower depth and poorer
resolution compared with
the HST survey so that they would include preferentially a larger number of  bright and large
clusters  than the HST surveys.
For example,
the blue LF for \citet{bed05} samples included in this study with $(B-V)_0 \leq 0.3$ has a peak
at the same magnitude to that for blue samples displayed in Fig.~\ref{lf}(a).
Thus it is natural that our LFs include a higher portion of faint star
clusters than those of \citet{cha01}.

\section{Summary and Conclusion}

We have carried out a survey of star clusters in M33  using the HST/WFPC2 images in the HST archive that
were not covered in the previous studies.
We have found 104 stars clusters including 32 new clusters and 12 known clusters
confirmed in this study from 24 fields of HST/WFPC2 images.
We combined 44 star clusters from this study, 154 star clusters from \citet{cha99a, cha01},
33 star clusters from \citet{bed05}, and 11 star clusters from
\citet{sar07a}, to build a catalog of 242 star clusters. This is the most comprehensive
catalog available to date which contains only the confirmed genuine star clusters in M33.
We present the integrated photometry of these star clusters derived from the $BVI$ images
taken with the CFH12K CCD camera at the CFHT. Photometric properties of the star clusters
are summarized as follows:

\begin{itemize}

\item The integrated color-magnitude diagram of star clusters shows that there are two distinct
populations in M33: a large number of blue star clusters and a small number of red star clusters. 
The CMD of the M33 star clusters is in general
similar to that of the LMC star clusters. 

\item We divide the cluster sample into three groups according to their $(B-V)_0$ color:
blue star clusters with $(B-V)_0\leq0.3$, 
intermediate color star clusters with $0.3<(B-V)_0<0.5$, and
red star clusters with $(B-V)_0\geq0.5$.
Most of the new clusters found or confirmed in this study are blue clusters.
There are 29 old globular cluster candidates with $0.5 \leq (B-V)_0 \leq 1.1$
and $0.7\leq(V-I)_0\leq1.2$, ten of which are newly found in this study.

\item Most of the new clusters in M33 are located along the theoretical evolutionary path 
for $Z=0.004$, $Y=0.24$ in $(B-V)_0$--$(V-I)_0$ color-color diagram, while a few of them
are in the redder side in the $(V-I)_0$ color.

\item The red clusters are found relatively more 
in the outer region of M33 than the blue and intermediate color clusters, and that
many of the blue stars are located in the HII regions.

\item The luminosity function (LF) for the blue star clusters shows a peak at 
$M_V \approx -7.3$ mag,  
while the intermediate color star cluster LF shows a peak at the fainter magnitude 
$M_V \approx -6.3$ mag. The LF for the red star clusters shows also a peak at $M_V \approx -6.8$,
although the number of the clusters is small.
\end{itemize}

\acknowledgments
Authors thank Narae Hwang 
for many fruitful discussions and comments during the work
in this study.
This work was supported in part supported by the ABRL(R14-2002-058-01000-0).
This is based in part on the W. -K. Park's Ph.D. thesis in Seoul National University.

\clearpage

\clearpage

\clearpage

\clearpage

\clearpage

\clearpage

\begin{deluxetable}{cccccccc}
\tabletypesize{\scriptsize}
\tablecaption{Summary of HST/WFPC2 fields used in this study \label{fld_list}}
\tablewidth{0pt}
\tablehead{
\colhead{} & \colhead{} & \multicolumn{4}{c}{Filters \& Exposure Times} &
\colhead{Number of Clusters} \\
\cline{3-6} \\
\colhead{ID} & \colhead{PropID} & \colhead{$U(F336W)$} & \colhead{$B(F439W)$} &
\colhead{$V(F505W)$} & \colhead{$I(F814W)$} & \colhead{new(all)}
}
\startdata
 1 & 6694 & $2\times800$s\tablenotemark{a} & $1\times500$s & $1\times100$s           &                             & 0(8)\\
 2 & 8805 &                        &              & $2\times700$s\tablenotemark{b} &                                & 2(6)\\
 3 & 8207 & $2\times400$s          & $2\times300$s & $2\times120$s                  &                               & 1(10)\\
 4 & 8207 & $2\times400$s          & $2\times300$s & $2\times120$s                  &                               & 3(7)\\
 5 & 8207 & $2\times400$s          & $2\times300$s & $2\times120$s                  &                               & 1(5)\\
 6 & 8805 &                        &              & $1\times1000$s\tablenotemark{b}, $1\times900$s\tablenotemark{b}& & 0(5)\\
 7 & 8018 &                        &              & $2\times1100$s, $2\times800$s   & $2\times1100$s                & 4(13)\\
 8 & 6431 &                        & $2\times350$s & $2\times260$s                  & $1\times260$s, $1\times200$s  & 5(13) \\
 9 & 9127 & $2\times260$s          & $2\times300$s & $2\times160$s                  &                               & 3(0) \\
10 & 8805 &                        &              & $2\times700$s\tablenotemark{b} &                                & 3(6)\\
11 & 8805 &                        &              & $2\times700$s\tablenotemark{b} &                                & 6(7)\\
12 & 7909 &                        &              & $2\times500$s\tablenotemark{b} &                                & 0(17)\\
13 & 8207 & $2\times400$s          & $2\times300$s & $2\times120$s                  &                               & 0(0)\\
14 & 9127 & $2\times260$s          & $2\times300$s & $2\times160$s                  &                               & 4(7)\\
15 & 8207 & $2\times400$s          & $2\times300$s & $2\times120$s                  &                               & 2(0)\\
16 & 5237 & $2\times600$s          &              & $2\times200$s                 & $2\times200$s                   & 0(0)\\
17 & 9127 & $2\times160$s, $2\times260$s & $2\times300$s & $2\times160$s                &                           & 2(0)\\
18 & 8207 & $2\times400$s           & $2\times300$s & $2\times120$s                 &                               & 1(1)\\
19 & 8061 & $1\times1600$s\tablenotemark{a} &      & $1\times700$s\tablenotemark{b} &                               & 1(1)\\
20 & 8207 & $2\times400$s           & $2\times300$s & $2\times120$s                  &                              & 0(0)\\
21 & 9479 &                        &              & $2\times140$s\tablenotemark{b} & $2\times500$s                  & 0(0)\\
22 & 9127 &                        &              & $2\times140$s\tablenotemark{b} & $2\times500$s                  & 0(0)\\
23 & 9127 &                        &              & $2\times140$s\tablenotemark{b} & $2\times500$s                  & 0(0)\\
24 & 8207 & $2\times400$s           & $2\times300$s & $2\times120s$                  &                              & 0(0)\\
\enddata
\tablenotetext{a}{Observed with F300W.}
\tablenotetext{b}{Observed with F606W.}
\end{deluxetable}

\begin{deluxetable}{rccrrrc}
\tabletypesize{\scriptsize}
\tablecaption{Catalog of new M33 star clusters discovered in this study\label{newcl_list}}
\tablewidth{0pt}
\tablehead{
\colhead{ID} & \colhead{RA(2000)} & \colhead{Dec(2000)} & \colhead{$V$\tablenotemark{a}} &
\colhead{$B-V$\tablenotemark{b}} & \colhead{$V-I$\tablenotemark{b}} & \colhead{$E(B-V)$\tablenotemark{c}}
}
\startdata
 1 & 1 32 36.36 & 30 36 48.61 & $  19.560\pm0.012 $ & $   0.130\pm0.007 $ & $   0.321\pm0.015 $ &   $ 0.20 $ \\ 
 2 & 1 33 11.74 & 30 38 56.14 & $  17.049\pm0.006\tablenotemark{\dagger} $ & $   0.092\pm0.008 $ & $   0.089\pm0.011 $ &   $ 0.10 $ \\ 
 3 & 1 33 12.67 & 30 38 55.35 & $  18.287\pm0.027 $ & $   0.231\pm0.037 $ & $   0.589\pm0.037 $ &   $ 0.10  $ \\ 
 4 & 1 33 14.97 & 30 39 04.70 & $  18.291\pm0.009 $ & $   0.479\pm0.009 $ & $   0.869\pm0.010 $ &   $ 0.15 $ \\ 
 5 & 1 33 19.51 & 30 35 30.19 & $  16.947\pm0.003 $ & $   0.156\pm0.002 $ & $   0.463\pm0.004 $ &   $ 0.10 $ \\ 
 6 & 1 33 20.25 & 30 34 50.27 & $  17.677\pm0.006 $ & $  -0.005\pm0.004 $ & $   0.111\pm0.009 $ &   $ 0.10 $ \\ 
 7 & 1 33 21.35 & 30 31 13.02 & $  18.719\pm0.013 $ & $   0.194\pm0.008 $ & $   0.482\pm0.012 $ &   $ 0.10 $ \\ 
 8 & 1 33 22.36 & 30 35 30.20 & $  20.144\pm0.056 $ & $   0.250\pm0.034 $ & $   0.696\pm0.052 $ &   $ 0.10 $ \\ 
 9 & 1 33 23.15 & 30 52 42.41 & $  19.220\pm0.011 $ & $   0.430\pm0.009 $ & $   0.654\pm0.010 $ &   $ 0.15 $ \\ 
10 & 1 33 26.58 & 30 31 30.82 & $  18.202\pm0.010 $ & $   0.258\pm0.007 $ & $   0.493\pm0.010 $ &   $ 0.10 $ \\ 
11 & 1 33 27.38 & 30 41 48.39 & $  19.017\pm0.025 $ & $   0.121\pm0.018 $ & $   0.426\pm0.032 $ &   $ 0.20 $ \\ 
12 & 1 33 30.06 & 30 31 48.66 & $  16.466\pm0.002 $ & $   0.110\pm0.003 $ & $   0.269\pm0.004 $ &   $ 0.10 $ \\ 
13 & 1 33 30.23 & 30 41 18.89 & $  18.489\pm0.018 $ & $   0.365\pm0.013 $ & $   0.695\pm0.016 $ &   $ 0.20 $ \\ 
14 & 1 33 30.36 & 30 37 43.41 & $  19.304\pm0.039 $ & $   0.319\pm0.034 $ & $   0.930\pm0.040 $ &   $ 0.15 $ \\ 
15 & 1 33 30.38 & 30 36 41.45 & $  19.034\pm0.031 $ & $   0.296\pm0.022 $ & $   0.907\pm0.032 $ &   $ 0.10 $ \\ 
16 & 1 33 30.41 & 30 36 00.04 & $  18.951\pm0.028 $ & $   0.361\pm0.021 $ & $   0.921\pm0.026 $ &   $ 0.10 $ \\ 
17 & 1 33 31.32 & 30 37 34.10 & $  19.798\pm0.069 $ & $   0.435\pm0.040 $ & $   1.180\pm0.043 $ &   $ 0.15 $ \\ 
18 & 1 33 32.16 & 30 30 18.01 & $  17.878\pm0.006 $ & $   0.238\pm0.005 $ & $   0.555\pm0.006 $ &   $ 0.10 $ \\ 
19 & 1 33 32.46 & 30 35 38.30 & $  16.308\pm0.004 $ &             \nodata & $   0.656\pm0.004 $ &   $ 0.15 $ \\ 
20 & 1 33 36.54 & 30 50 35.79 & $  17.574\pm0.004 $ & $   0.038\pm0.004 $ & $  -0.082\pm0.009 $ &   $ 0.15 $ \\ 
21\tablenotemark{*} & 1 33 38.19 & 30 43 24.51 & $  18.570\pm0.012 $ & $   0.739\pm0.026 $ &     \nodata &   $ 0.15 $ \\ 
22 & 1 33 41.44 & 30 51 10.51 & $  17.934\pm0.058 $ & $   0.448\pm0.057 $ & $   1.085\pm0.076 $ &   $ 0.15  $ \\ 
23 & 1 33 58.09 & 30 37 29.68 & $  17.771\pm0.013 $ & $   0.137\pm0.015 $ & $   0.578\pm0.022 $ &   $ 0.20 $ \\ 
24 & 1 33 58.49 & 30 38 20.96 & $  16.790\pm0.005 $ & $   0.097\pm0.006 $ & $   0.454\pm0.013 $ &   $ 0.20 $ \\ 
25 & 1 33 58.67 & 30 35 26.10 & $  16.156\pm0.007 $ & $  -0.045\pm0.005 $ & $   0.012\pm0.009 $ &   $ 0.10 $ \\ 
26 & 1 34 14.44 & 30 34 17.56 & $  16.879\pm0.013 $ & $   0.097\pm0.013 $ & $   0.629\pm0.016 $ &   $ 0.10 $ \\ 
27 & 1 34 14.96 & 30 33 53.66 & $  17.864\pm0.013 $ & $   0.035\pm0.011 $ & $   0.398\pm0.015 $ &   $ 0.10 $ \\ 
28 & 1 34 15.21 & 30 33 15.11 & $  18.512\pm0.014 $ & $   0.660\pm0.014 $ & $   1.047\pm0.012 $ &   $ 0.10 $ \\ 
29 & 1 34 15.74 & 30 33 41.02 & $  16.834\pm0.011 $ & $   0.172\pm0.011 $ & $   0.765\pm0.012 $ &   $ 0.10 $ \\ 
30 & 1 34 18.08 & 30 31 58.93 & $  18.154\pm0.006 $ & $  -0.016\pm0.005 $ & $   0.064\pm0.012 $ &   $ 0.10 $ \\ 
31 & 1 34 42.79 & 30 37 20.21 & $  17.652\pm0.003 $ & $   0.197\pm0.002 $ & $   0.810\pm0.003 $ &   $ 0.10 $ \\ 
32 & 1 34 44.60 & 30 37 32.32 & $  18.071\pm0.005 $ & $   0.023\pm0.009 $ & $   0.166\pm0.015 $ &   $ 0.05 $ \\ 
\enddata
\tablenotetext{a}{Measured on CFH12k CCD image with $r=4.0\arcsec$ aperture.}
\tablenotetext{b}{Measured on CFH12k CCD image with $r=2.0\arcsec$ aperture.}
\tablenotetext{c}{Measured with $UBVRI$ PSF Photometry data by \citet{mas06}. The uncertainties of the values are 
$0.05$.}
\tablenotetext{\dagger}{Happened to be very close to the edge of the CCD chip. So the measurement was made with
smaller apertures.}
\tablenotetext{*}{Measured on HST/WFPC2 image instead of CFH12k CCD image, with $r=2.2\arcsec$ and $r=1.0\arcsec$ 
     apertures for $V$ and color measurement, respectively.}
\end{deluxetable}

\begin{deluxetable}{rccrrrcl}
\tabletypesize{\scriptsize}
\tablecaption{Catalog of previously known M33 star clusters confirmed in this study\label{othercl_list}}
\tablewidth{0pt}
\tablehead{
\colhead{ID} & \colhead{RA(2000)} & \colhead{Dec(2000)} & \colhead{$V$\tablenotemark{a}} &
\colhead{$B-V$\tablenotemark{b}} & \colhead{$V-I$\tablenotemark{b}} &
\colhead{$E(B-V)$\tablenotemark{c}} & \colhead{Cross Identification\tablenotemark{d}}
}
\startdata
 33 & 1 32 34.45 & 30 37 42.09 & $  20.180\pm0.019 $ & $   0.559\pm0.016 $ & $   0.725\pm0.020 $ &   $ 0.05 $ & CBF-143 \\ 
 34 & 1 33 22.12 & 30 40 26.15 & $  18.090\pm0.010 $ & $   0.080\pm0.007 $ & $   0.360\pm0.014 $ &   $ 0.10 $ & CBF-59 \\ 
 35 & 1 33 22.26 & 30 40 59.64 & $  18.663\pm0.017 $ &    \nodata\tablenotemark{*} & $   0.475\pm0.018 $ &   $ 0.10 $ & CBF-60 \\ 
 36 & 1 33 22.35 & 30 30 14.67 & $  17.378\pm0.004 $ & $   0.130\pm0.002 $ & $   0.460\pm0.004 $ &   $ 0.10 $ & CS-H36 \\ 
 37 & 1 33 23.84 & 30 40 26.18 & $  19.480\pm0.040 $ & $   0.177\pm0.026 $ & $   0.685\pm0.041 $ &   $ 0.10 $ & CBF-96 \\ 
 38 & 1 33 24.81 & 30 33 55.25 & $  20.429\pm0.077 $ & $  -0.117\pm0.032 $ & $   0.358\pm0.068 $ &   $ 0.10 $ & CBF-18 \\ 
 39 & 1 33 25.60 & 30 29 57.03 & $  18.179\pm0.006 $ & $   0.852\pm0.005 $ & $   1.115\pm0.004 $ &   $ 0.10 $ & CS-U126, MKKSS-5 \\ 
 40 & 1 33 25.96 & 30 36 24.58 & $  17.525\pm0.007 $ & $   0.265\pm0.005 $ & $   0.610\pm0.007 $ &   $ 0.10 $ & MKKSS-6 \\ 
 41 & 1 33 26.34 & 30 41 06.96 & $  18.909\pm0.021 $ & $   0.477\pm0.014 $ & $   1.043\pm0.016 $ &   $ 0.20 $ & CBF-92 \\ 
 42 & 1 33 26.46 & 30 41 11.84 & $  19.358\pm0.028 $ & $   0.361\pm0.022 $ & $   0.786\pm0.031 $ &   $ 0.20 $ & CBF-93 \\ 
\enddata
\tablenotetext{a}{Measured on CFH12k CCD image with $r=4.0''$ aperture.}
\tablenotetext{b}{Measured on CFH12k CCD image with $r=2.0''$ aperture.}
\tablenotetext{c}{Measured with $UBVRI$ PSF Photometry data by \citet{mas06}. The uncertainties of the values are 
$0.05$.}
\tablenotetext{d}{B identifications are from \citet{bed05}; CBF identifications from \citet{cha99a, cha01};
CS identification from \citet{cs82}; MKKSS identifications from \citet{moc98}; MD identifications are
from \citet{md78} based on the coordinates given by \citet{ma02} ; and S identifications from \citet{sar07a}.}
\tablenotetext{*}{Happened to be off from CFHT CFH12k images.}
\tablecomments{The complete version of this table is in the electronic edition of
the Journal. The printed edition contains only a sample.}
\end{deluxetable}

\begin{deluxetable}{rccrrrcl}
\tabletypesize{\scriptsize}
\tablecaption{Catalog of previously known M33 star clusters\label{prevcl_list}}
\tablewidth{0pt}
\tablehead{
\colhead{ID} & \colhead{RA(2000)} & \colhead{Dec(2000)} & \colhead{$V$\tablenotemark{a}} &
\colhead{$B-V$\tablenotemark{b}} & \colhead{$V-I$\tablenotemark{b}} &
\colhead{$E(B-V)$\tablenotemark{c}} & \colhead{Cross Identification\tablenotemark{d}}
}
\startdata
105\tablenotemark{*} & 1 32 38.83 & 30 39 17.77 & $19.942\pm0.065$ & $ 0.267\pm0.090$ &    \nodata &   $ 0.10 $ & CBF-162 \\ 
106 & 1 32 44.28 & 30 40 12.25 & $  18.434\pm0.005 $ & $   0.749\pm0.006 $ & $   1.065\pm0.006 $ &   $ 0.10 $ & CBF-161, CS-U88 \\ 
107 & 1 33 01.16 & 30 35 44.65 & $  19.485\pm0.026 $ & $   0.449\pm0.029 $ & $   1.152\pm0.028 $ &   $ 0.05 $ & CBF-39 \\ 
108 & 1 33 02.39 & 30 34 44.16 & $  18.022\pm0.006 $ & $  -0.034\pm0.004 $ & $   0.082\pm0.011 $ &   $ 0.15 $ & CBF-40 \\ 
109 & 1 33 08.10 & 30 28 00.12 & $  19.062\pm0.015 $ & $   0.638\pm0.011 $ & $   0.822\pm0.013 $ &   $ 0.15 $ & CBF-86, CS-U140 \\ 
110 & 1 33 10.08 & 30 29 56.51 & $  18.523\pm0.022 $ & $   0.296\pm0.016 $ & $   0.483\pm0.018 $ &   $ 0.15 $ & CBF-89 \\ 
111 & 1 33 13.84 & 30 29 05.27 & $  18.884\pm0.013 $ & $   0.837\pm0.014 $ & $   0.999\pm0.015 $ &   $ 0.20 $ & CBF-53 \\ 
112 & 1 33 13.88 & 30 29 44.87 & $  17.929\pm0.012 $ & $   0.042\pm0.013 $ & $  -0.069\pm0.018 $ &   $ 0.15 $ & CBF-88 \\ 
113 & 1 33 14.28 & 30 28 22.78 & $  18.298\pm0.006 $ & $   0.874\pm0.005 $ & $   1.109\pm0.004 $ &   $ 0.10 $ & CBF-54, CS-U137, MD-8 \\ 
114 & 1 33 15.15 & 30 32 53.26 & $  19.714\pm0.026 $ & $   0.629\pm0.018 $ & $   0.856\pm0.018 $ &   $ 0.05 $ & CBF-144 \\ 
\enddata
\tablenotetext{a}{Measured on CFH12k CCD image with $r=4.0''$ aperture.}
\tablenotetext{b}{Measured on CFH12k CCD image with $r=2.0''$ aperture.}
\tablenotetext{c}{Measured with $UBVRI$ PSF Photometry data by \citet{mas06}. The uncertainties of the values are 
$0.05$.}
\tablenotetext{d}{B identifications are from \citet{bed05}; CBF identifications from \citet{cha99a, cha01};
CS identification from \citet{cs82}; MKKSS identifications from \citet{moc98}; MD identifications are
from \citet{md78} based on the coordinates given by \citet{ma02} ; and S identifications from \citet{sar07a}.}
\tablenotetext{*}{Measured on HST/WFPC2 image instead of CFH12k CCD image, with $r=2.2\arcsec$ and $r=1.0\arcsec$ 
     apertures for $V$ and color measurement, respectively.}
\tablecomments{The complete version of this table is in the electronic edition of
the Journal. The printed edition contains only a sample.}
\end{deluxetable}

\begin{deluxetable}{rrrrcl}
 \tabletypesize{\scriptsize}
\tablecaption{List of globular cluster candidates\label{gccand_list}}
\tablewidth{0pt}
\tablehead{
\colhead{ID} & \colhead{$V$\tablenotemark{a}} & \colhead{$B-V$\tablenotemark{b}} & \colhead{$V-I$\tablenotemark{b}}&
\colhead{$E(B-V)$\tablenotemark{c}} & \colhead{Cross Identification\tablenotemark{d}}
}
\startdata
 28 & $  18.512\pm0.014 $ & $   0.660\pm0.014 $ & $   1.047\pm0.012 $ &   $ 0.10 $ & \\ 
 39 & $  18.179\pm0.006 $ & $   0.852\pm0.005 $ & $   1.115\pm0.004 $ &   $ 0.10 $ & CS-U126, MKKSS-5 \\ 
 45 & $  18.348\pm0.011 $ & $   0.789\pm0.009 $ & $   1.140\pm0.009 $ &   $ 0.20 $ & CBF-90, CS-U77, MD-17 \\ 
 49 & $  17.322\pm0.006 $ & $   0.759\pm0.005 $ & $   0.976\pm0.004 $ &   $ 0.20 $ & CBF-90 \\ 
 58 & $  19.497\pm0.011 $ & $   0.885\pm0.015 $ & $   1.098\pm0.015 $ &   $ 0.25 $ & SBGHS-10 \\ 
 76 & $  17.889\pm0.016 $ & $   0.711\pm0.018 $ & $   0.942\pm0.013 $ &   $ 0.10 $ & CBF-49, MKKSS-25 \\ 
 84 & $  18.454\pm0.034 $ & $   0.742\pm0.050 $ & $   1.185\pm0.040 $ &   $ 0.15 $ & CBF-29 \\ 
 93 & $  17.791\pm0.009 $ & $   0.658\pm0.009 $ & $   0.920\pm0.011 $ &   $ 0.10 $ & CBF-118 \\ 
 94 & $  18.155\pm0.011 $ & $   0.600\pm0.011 $ & $   0.888\pm0.014 $ &   $ 0.10 $ & CBF-119 \\ 
106 & $  18.434\pm0.005 $ & $   0.749\pm0.006 $ & $   1.065\pm0.006 $ &   $ 0.10 $ & CBF-161, CS-U88 \\ 
111 & $  18.884\pm0.013 $ & $   0.837\pm0.014 $ & $   0.999\pm0.015 $ &   $ 0.20 $ & CBF-53 \\ 
113 & $  18.298\pm0.006 $ & $   0.874\pm0.005 $ & $   1.109\pm0.004 $ &   $ 0.10 $ & CBF-54, CS-U137, MD-8 \\ 
114 & $  19.714\pm0.026 $ & $   0.629\pm0.018 $ & $   0.856\pm0.018 $ &   $ 0.05 $ & CBF-144 \\ 
130 & $  19.012\pm0.013 $ & $   0.685\pm0.013 $ & $   1.043\pm0.013 $ &   $ 0.15 $ & CBF-108, SBGHS-6 \\ 
131 & $  18.215\pm0.006 $ & $   1.244\pm0.007 $ & $   1.291\pm0.005 $ &   $ 0.20 $ & CBF-106, SBGHS-4 \\ 
146 & $  16.290\pm0.002 $ & $   0.645\pm0.001 $ & $   0.993\pm0.001 $ &   $ 0.10 $ & CBF-61, CS-U49, MD-24, MKKSS-21 \\ 
149 & $  19.042\pm0.021 $ & $   0.606\pm0.016 $ & $   0.967\pm0.013 $ &   $ 0.10 $ & CBF-62 \\ 
157 & $  19.878\pm0.053 $ & $   0.605\pm0.063 $ & $   0.920\pm0.068 $ &   $ 0.10 $ & CBF-138 \\ 
163 & $  17.261\pm0.003 $ & $   0.784\pm0.003 $ & $   1.082\pm0.003 $ &   $ 0.10 $ & CBF-104, CS-H38, MD-25, MKKSS-22 \\ 
189 & $  16.390\pm0.003 $ & $   0.707\pm0.002 $ & $   0.973\pm0.002 $ &   $ 0.15 $ & CBF-28 \\ 
191 & $  16.512\pm0.008 $ & $   0.929\pm0.007 $ & $   1.298\pm0.005 $ &   $ 0.10 $ & CBF-98, MKKSS-31 \\ 
196 & $  19.149\pm0.018 $ & $   0.844\pm0.029 $ & $   1.266\pm0.022 $ &   $ 0.15 $ & B-5 \\ 
206 & $  16.354\pm0.008 $ & $   0.947\pm0.007 $ & $   1.160\pm0.006 $ &   $ 0.10 $ & CBF-116, MKKSS-36 \\ 
214 & $  18.354\pm0.013 $ & $   0.621\pm0.013 $ & $   1.106\pm0.013 $ &   $ 0.10 $ & CBF-11 \\ 
228 & $  17.123\pm0.003 $ & $   0.685\pm0.002 $ & $   0.989\pm0.001 $ &   $ 0.10 $ & CBF-70, CS-C03, MD-47, SBGHS-13 \\ 
234 & $  19.677\pm0.025 $ & $   0.636\pm0.024 $ & $   1.066\pm0.022 $ &   $ 0.10 $ & CBF-75 \\ 
236 & $  18.793\pm0.008 $ & $   0.827\pm0.007 $ & $   1.030\pm0.006 $ &   $ 0.10 $ & CBF-74, CS-H24 \\ 
240 & $  20.921\pm0.060 $ & $   0.906\pm0.048 $ & $   1.366\pm0.038 $ &   $ 0.15 $ & CBF-82 \\ 
241 & $  17.550\pm0.003 $ & $   0.821\pm0.004 $ & $   1.032\pm0.003 $ &   $ 0.15 $ & CBF-81, CS-C20, MD-55 \\ 
\enddata
\tablenotetext{a}{Measured on CFH12k CCD image with $r=4.0''$ aperture.}
\tablenotetext{b}{Measured on CFH12k CCD image with $r=2.0''$ aperture.}
\tablenotetext{c}{Measured with $UBVRI$ PSF Photometry data by \citet{mas06}. The uncertainties of the values are 
$0.05$.}
\tablenotetext{d}{B identifications are from \citet{bed05}; CBF identifications from \citet{cha99a, cha01};
CS identification from \citet{cs82}; MKKSS identifications from \citet{moc98}; MD identifications are
from \citet{md78} based on the coordinates given by \citet{ma02} ; and S identifications from \citet{sar07a}.}
\end{deluxetable}


\begin{thebibliography}{}
\bibitem[Bedin \etal (2005)]{bed05} Bedin, L. R., Piotto, G., Baume, G., Momany, Y., Carraro, G.,
         Anderson, J., Messineo, M., \& Ortolani, S. 2005, \aap, 444, 831
\bibitem[Bica \etal (1996)]{bic96} Bica, E., Claria, J. J., Dottori, H., Santos, J. F. C., Jr.,
        \&  Piatti, A. E. 1996, \apjs, 102, 57
\bibitem[Bruzual \& Charlot (2003)]{bru03} Bruzual, G., \& Charlot, S. 2003, \mnras, 344, 1000
\bibitem[Chandar \etal (1999a)]{cha99a} Chandar, R., Bianchi, L., \& Ford, H. C. 1999a, 
        \apjs, 122, 431
\bibitem[Chandar \etal (1999b)]{cha99b} Chandar, R., Bianchi, L., \& Ford, H. C. 1999b, 
        \apj, 517, 668
\bibitem[Chandar \etal (2001)]{cha01} Chandar, R., Bianchi, L., \& Ford, H. C. 2001, 
        \aap, 366, 498
\bibitem[Christian \& Schommer (1982)]{cs82} Christian, C. A., \& Schommer, R. A. 1982, 
        \apjs, 49, 405
\bibitem[Christian \& Schommer (1988)]{cs88} Christian, C. A., \& Schommer, R. A. 1988, 
        \aj, 95, 704
\bibitem[Corbelli (2003)]{cor03} Corbelli, E. 2003, \mnras, 342, 199
\bibitem[Harris (1996)]{har96} Harris, W. E. 1996, \aj, 112, 1487
\bibitem[Hiltner (1960)]{hil60} Hiltner, W. A. 1960, \apj, 131, 163
\bibitem[Hodge \etal (1999)]{hod99} Hodge, P. W., Balsley, J., Wyder, T. K., \& Skelton, B. P. 1999,   \pasp, 111, 685
\bibitem[Hwang \& Lee (2007)]{hwa07} Hwang, N., \& Lee, M. G. 2007, \aj, submitted
\bibitem[Kaluzny \etal (1998)]{kal98} Kaluzny, J., Stanek, K. Z., Krockenberger, M., Sasselov, D. D.,
        Tonry, J. L., \& Mateo, M. 1998, \aj, 115, 1016 
\bibitem[Kim et al.(2002)]{kim02} Kim, M., Kim, E., Lee, M. G., Sarajedini, A.,
        \& Geisler, D. 2002, \aj, 123, 244
\bibitem[Kron \& Mayall (1960)]{km60} Kron, G. E., \& Mayall, N. U. 1960, \aj, 65, 581
\bibitem[Lata \etal (2002)]{lat02} Lata, S., Pandey, A. K., Sagar, R., \& Mohan, V. 2002, \aap,
        388, 158
\bibitem[Lee (2006)]{lee06} Lee, M. G. 2006, Bull. Astr. Soc. India, 34, 99
\bibitem[Ma \etal (2002)]{ma02} Ma, J., Zhou, X., Chen, J. -S., Wu, H., Jiang, Z. -J., Xue, S. -J., 
        \& Zhu, J. 2002, \cjaa, 2, 197
\bibitem[Massey et al.(1995)]{mas95} Massey, P., Armandroff, T. E., Pyke, R., Patel, K.,
        \& Wilson, C. 1995, \aj, 110, 2715
\bibitem[Massey \etal (2006)]{mas06} Massey, P., Olsen, K. A. G., Hodge, P. W., Strong, S. B.,
        Jacoby, G. H., Schlingman, W., \& Smith, R. C. 2006, \aj, 131, 2478
\bibitem[Melnick \& D'Odorico (1978)]{md78} Melnick, J., \& D'Odorico, S. 1978, \aaps, 34, 249
\bibitem[Mochejska \etal (1998)]{moc98} Mochejska, B. J., Kaluzny, J., Krockenberger, M.,
        Sasselov, D. D., \& Stanek, K. Z. 1998, \actaa, 48, 455
\bibitem[Sarajedini \etal (2000)]{sar00} Sarajedini, A., Geisler, D.,  Schommer, R.,
      \& Harding, P. 2000, \aj, 120, 2437
\bibitem[Sarajedini \etal (2007)]{sar07a} Sarajedini, A., Barker, M. K., Geisler, D., Harding, P., 
      \& Schommer, R. 2007, \aj, 133, 290
\bibitem[Sarajedini \& Mancone (2007)]{sar07b} Sarajedini, A., \& Mancone, C. L. 2007, \aj, 134, 447
\bibitem[Schlegel \etal (1998)]{sch98} Schlegel, D. J., Finkbeiner, D. P., \& Davis, M. 1998,
      \apj, 500, 525
\bibitem[Stanek \etal (1998)]{sta98} Stanek, K. Z., Kaluzny, J., Krockenberger, M., Sasselov, D. D.,
        Tonry, J. L., \& Mateo, M. 1998, \aj, 115, 1894
\bibitem[Zaritsky \etal (1989)]{zar89} Zaritsky, D., Elston, R., \& Hill, J. M. 1989,
      \aj, 97, 97
\end{thebibliography}
\end{document}